\documentclass[a4paper]{PoS}

\usepackage{mathptm}  
\usepackage{amsmath} 
\usepackage{epsfig}
\usepackage{subfig} 

\usepackage{ulem} 
\usepackage[babel=true]{csquotes}    

\newcommand{\beq}{\begin{eqnarray}}
\newcommand{\eeq}{\end{eqnarray}}
\newcommand{\be}{\begin{eqnarray*}}
\newcommand{\ee}{\end{eqnarray*}}

\newcommand{\ie}{{\it i.e.}}

\newcommand{\etal}{{\it et al.}}

\newcommand{\cf}[1]{{Fig.~\ref{#1}}}
\newcommand{\ct}[1]{{Table~\ref{#1}}}

\def\lsim{\raise0.3ex\hbox{$<$\kern-0.75em\raise-1.1ex\hbox{$\sim$}}}
\def\gsim{\raise0.3ex\hbox{$>$\kern-0.75em\raise-1.1ex\hbox{$\sim$}}}

\def\dAu  {$d$Au}
\def\dAum  {d\mathrm{Au}}

\def\pp   {$pp$}
\def\pA   {$pA$}
\def\AA   {$AA$}

\def\pT      {\mbox{$P_{T}$}}

\def\beq     {\begin{equation}}
\def\eeq     {\end{equation}}

\newcommand{\QQ}{\mbox{$Q\bar{Q}$}}
\newcommand{\cc}{\mbox{$c\bar{c}$}}
\def\jpsi    {\mbox{$J/\psi$}}

\newcommand{\upsi}{\mbox{$\Upsilon$}}
\newcommand{\bb}{\mbox{$b\bar{b}$}}

\title{Cold Nuclear Matter effects in \upsi\ production in \dAu\ collisions at RHIC}

\ShortTitle{CNM effects in \upsi\ production in \dAu\ collisions at RHIC}

\author{\speaker{A.~Rakotozafindrabe}\\
        IRFU/SPhN, CEA Saclay, 91191 Gif-sur-Yvette Cedex, France
        }

\author{E.G.~Ferreiro\\
        DFP \& IGFAE, Universidade de Santiago de Compostela, 15782 Santiago de Compostela, Spain
        }

\author{F.~Fleuret\\
        Laboratoire Leprince Ringuet, \'Ecole Polytechnique, CNRS/IN2P3,  91128 Palaiseau, France
        }

\author{J.P.~Lansberg\\
        IPNO, Universit\'e Paris-Sud, CNRS/IN2P3, F-91406, Orsay, France
}

\author{N.~Matagne\\
        Universit\'e de Mons, Service de Physique Nucl\'eaire et Subnucl\'eaire, Place du Parc 20, B-7000 Mons, Belgium
}

\abstract{
We report on our recent study of Cold Nuclear Matter effects on the \upsi\ production at RHIC in $d$Au collisions.
The first experimental results available on the nuclear modification factor $R^{\Upsilon}_{\dAum}$ have rather large uncertainties. They nevertheless allow to bring qualitative information on the nature of the 
nuclear effects at play on top of the usual nuclear absorption, since the latter is expected to lie in a quite small range 
around a value close to ten times smaller as for
charmonia. At backward rapidities, the behavior of $R^{\Upsilon}_{d\rm Au}$ hints at the presence of a 
gluon EMC effect, analogous to the quark EMC effect -- but possibly stronger. Mid rapidity measurements 
with a better precision are highly desirable to pin down the gluon anti-shadowing, still under debate. 
At forward rapidities, the data leave some room for an additional  
{\it fractional} energy loss mechanism, recently revived in the literature.}

\FullConference{Sixth International Conference on Quarks and Nuclear Physics\\
		 April 16-20, 2012\\
		 Ecole Polytechnique, Palaiseau,  Paris}
		 		 
\begin{document}


\section{Introduction}

Heavy-quarkonium (\QQ) in-vacuum studies in proton-proton collisions allow to probe Quantum Chromodynamics (QCD)
at short and long distances via their production mechanisms~\cite{Lansberg:2006dh,Brambilla:2010cs}. In-medium studies~\cite{Brambilla:2010cs,Rapp:2008tf} 
give access to Cold Nuclear Matter (CNM) effects in proton-nucleus collisions and to the understanding 
of QCD at high density and temperature in nucleus-nucleus collisions. STAR and PHENIX experiments~\cite{Reed:2010zzb} 
at RHIC provided the first measurements of the rapidity dependence of the inclusive $\Upsilon(b\bar{b})$ production in \dAu\ collisions at 
$\sqrt{s_{{NN}}}=200 \,\mathrm{GeV}$. They are expressed in terms of the nuclear modification factor $R^{\Upsilon}_{\dAum}$ of the \upsi~yield obtained in \dAu\ collisions with respect to the superposition of the equivalent number of \pp\ collisions at the same energy. Despite the rather large uncertainties on $R^{\Upsilon}_{\dAum}$, we have shown~\cite{Ferreiro:2011xy} that it still gives qualitative insights into the nature of the CNM effects at play for \upsi\ production in \dAu\ collisions.

At high energy (small Bjorken-$x$), the nucleus in the rest frame of the incoming proton is Lorentz-contracted and the nucleon wave functions 
within this nucleus overlap. As a consequence, due to non-linear effects caused by the interactions of these overlapping nucleons, 
the nuclear Parton Distribution Functions (nPDF) deviate from those of free nucleons.  Nucleons {\it shadow} \cite{Glauber:1955qq,Gribov:1968jf} 
each other and one expects the nPDFs to show lower values than for free nucleons.  For energies corresponding to $0.01 \leq x_B \leq 0.3$, some 
experimental data point~\cite{Gousset:1996xt} at an excess of partons compared to free nucleons, dubbed as anti-shadowing.  

In the following, we report on our recent study~\cite{Ferreiro:2011xy} where we have showed that usual nuclear modifications of gluon distribution in 
heavy ions -- the shadowing and the anti-shadowing -- as well as the possible break up (nuclear absorption) of the \bb\ pair 
when it passes through the nucleus have a small effect: the resulting description of both the overall trend and magnitude
 of the \upsi\ data in \dAu\ collisions appears rather unsatisfactory. Let us note that the mid rapidity measurement corresponds to the anti-shadowing region.

Interestingly, we have pointed out that the modification of the \upsi\ yield seen at backward rapidity corresponds to 
the intermediate Bjorken-$x$ region, $0.35 \leq x_B \leq 0.7$, where a further nuclear suppression of the gluon distribution 
could be expected but was unobserved until now. Such an effect would be analogous to the one reported by the EMC
 collaboration~\cite{Aubert:1983rq} for the quarks. To date, there is no consensus \cite{Norton:2003cb} to explain this suppression, 
known as the (quark) EMC effect. It is still the object of intense investigations. 
This effect has not yet been clearly observed for gluons,
even though it is allowed in some fits of gluon nPDFs. The gluon EMC suppression  is usually overlooked 
whereas the shadowing of gluons is the subject of intense on-going discussions.
Not much is known about gluons in this region and few data 
are sensitive to their distribution at $x_B$ larger than 0.3. The amount of the EMC suppression is 
actually basically unknown~\cite{Eskola:2009uj}, except for a loose constrain set by 
the momentum sum rule. We explore the impact on $R^{\Upsilon}_{\dAum}$ that results from the use of the moderate and limiting cases 
currently offered by the nPDF parametrisations in this $x_B$ region. 
By comparing our results to the RHIC data, we can conclude that this provides a first and interesting hint for a strong gluon EMC effect, 
which might be stronger than the quark one.

As regards the forward rapidity data, the discrepancy with the usual CNM effects cited above -- shadowing and nuclear absorption -- 
points at the presence of an additional CNM effect: a {\it fractional} energy loss~\cite{Arleo:2010rb}
proportional to the projectile {parton} energy {and} caused by medium-induced radiations 
associated to the quarkonium hadroproduction. This radiative energy loss arises when the incoming parton and the outgoing 
coloured object radiate nearly coherently. This is possible when the colour-charge flow  is subject to a scattering at a small angle 
in the nucleus rest frame. The bound on parton energy loss discussed previously in ~\cite{Brodsky:1992nq},
which forbids energy loss to scale with energy, does not apply for such radiations. This {\it fractional} energy loss is 
thus probably at work at RHIC energies, 
contrary to the usual radiative energy loss~\cite{Baier:1996sk}. Such an effect might induce an extra suppression of about 10 to $20\%$ of the \upsi\ yield in \dAu\ at forward rapidity at RHIC. 

In practice, for the \upsi\ production, we have used our well established Monte-Carlo framework {\sf JIN}~\cite{Ferreiro:2008qj} -- based on the probabilistic Glauber model and used to describe \jpsi\ production at RHIC -- with the following ingredients: the partonic process for the \bb\ production and the CNM effects. For now, the three \upsi\ resonances are 
not resolved but are measured together. Since the nuclear absorption has 
to be small (as discussed in Section~\ref{sec:eff-sigma-abs}) and since the nPDF effects are very likely similar for these three 
states, we have safely considered them on the same footage.

\vspace*{-0.2cm}

\section{Partonic process for the \bb\ production}
\label{sec:partonic-process}

We have considered improved kinematics corresponding to a $2\to2$ ($g+g\rightarrow b \bar{b}+g$) partonic process for the \upsi\ production. 
In earlier studies of CNM effects on \upsi\ production~\cite{Vogt:2010aa}, the $b \bar b$ pair has been assumed to be the result of a $2\to 1$ 
partonic process (\ie~$g+g\rightarrow b \bar{b}$). The presence of a final-state gluon introduces further degrees
of freedom in the kinematics, allowing several $(x_1\, , \,x_2)$ for a given set $(y\, , \,P_T)$ with the measured \upsi\ rapidity and transverse momentum.
Kinematics determines
the physical phase space, but models are {\it mandatory} to compute the proper
weighting of each kinematically allowed $(x_1, x_2)$. This weight is simply
the differential cross section at the partonic level times the gluon PDFs,
\ie\ $g(x_1,\mu_F) g(x_2, \mu_F) \, d\sigma_{gg\to \Upsilon + g} /dy \,
dP_T\, dx_1 dx_2 $.
In the present status of our code, we are able to use the partonic differential
cross section computed from {\it any} approach. For the present study, we have used the Colour-Singlet Model (CSM) at LO~\cite{Brodsky:2009cf}, which offers a good description of the direct $\Upsilon{\mathrm(1S)}$ (see \cf{fig:CSM-LO}) and direct $\Upsilon{\mathrm(3S)}$ production at low \pT\ (where lies the bulk of the integrated cross-section). 

\vspace*{-0.2cm}

\section{Gluon-momentum distribution in nuclei}
\label{sec:gluon-nPDF}

To obtain the \upsi\ yield in \pA\ and \AA\ collisions, a correction
factor due to the nuclear modification of the gluon-momentum distribution has to be applied to the \upsi\ yield obtained from the simple
superposition of the equivalent number of \pp\ collisions.
This factor can be expressed in terms of the ratios $R_i^A$ 
of the nPDF of a nucleon bound in a nucleus~$A$ to the free nucleon PDF. The 
numerical parametrisation of $R_i^A(x_B,Q^2)$ is given for each parton 
flavours. We have restricted our study to gluons since, at RHIC, \upsi\ is essentially 
produced through the fusion of gluons~\cite{Lansberg:2006dh,Brambilla:2010cs}. As usually done, we label $x_1$ ($x_2$)
the gluon momentum fraction in the proton/deuteron (nucleus). The nPDF spatial dependence has been included with a modification 
proportional to the local density~\cite{Klein:2003dj}.

\begin{figure}[htb!]
\begin{center}
%
\subfloat[][Data vs CSM LO for the direct \upsi(1S) cross-section.]{%
\label{fig:CSM-LO}
\includegraphics[width=.49\linewidth]{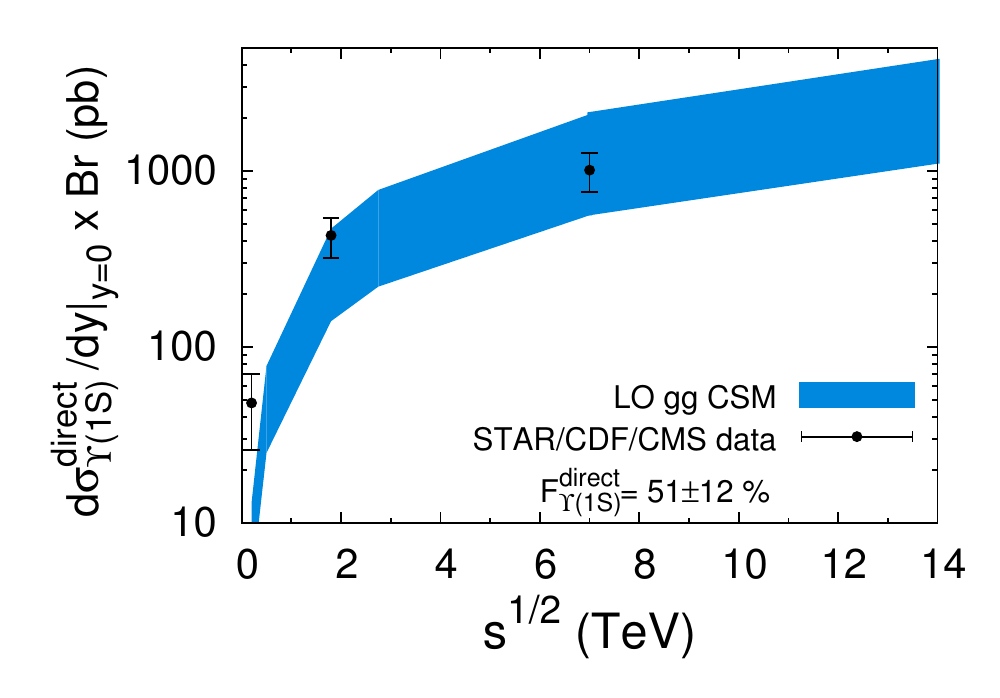}}
%
%
\subfloat[][(Anti-)shadowing and $\sigma_{\mathrm{eff}}$ uncertainties on $R^{\Upsilon}_{\dAum}$.]{%
\label{fig:EMC-nPDF-sigma-b}
\includegraphics[width=.49\linewidth]{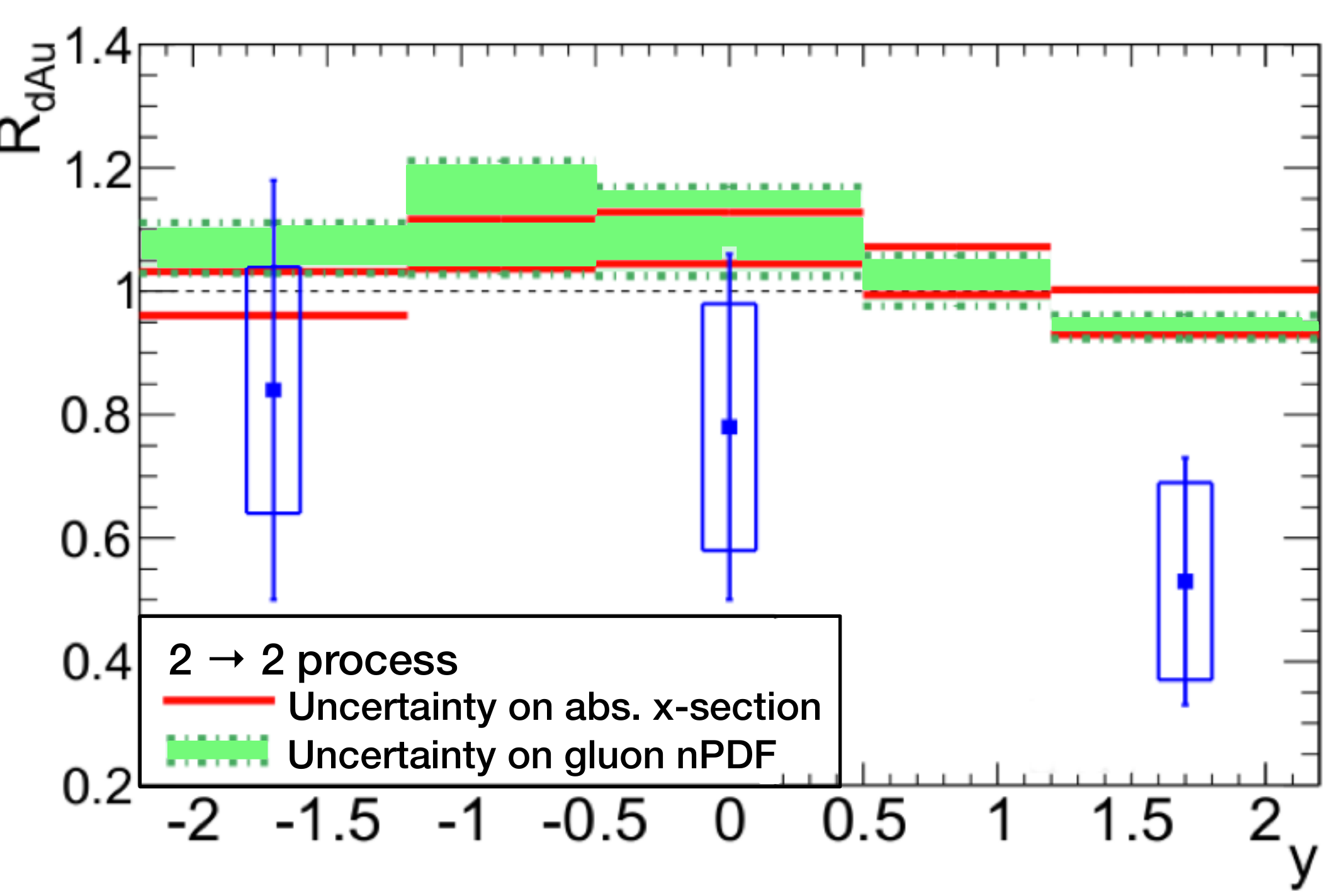}} \\
%
\subfloat[][EPS09 limiting curves in the EMC region.]{%
\label{fig:EMC-nPDF-sigma-c}
\includegraphics[width=.30\linewidth]{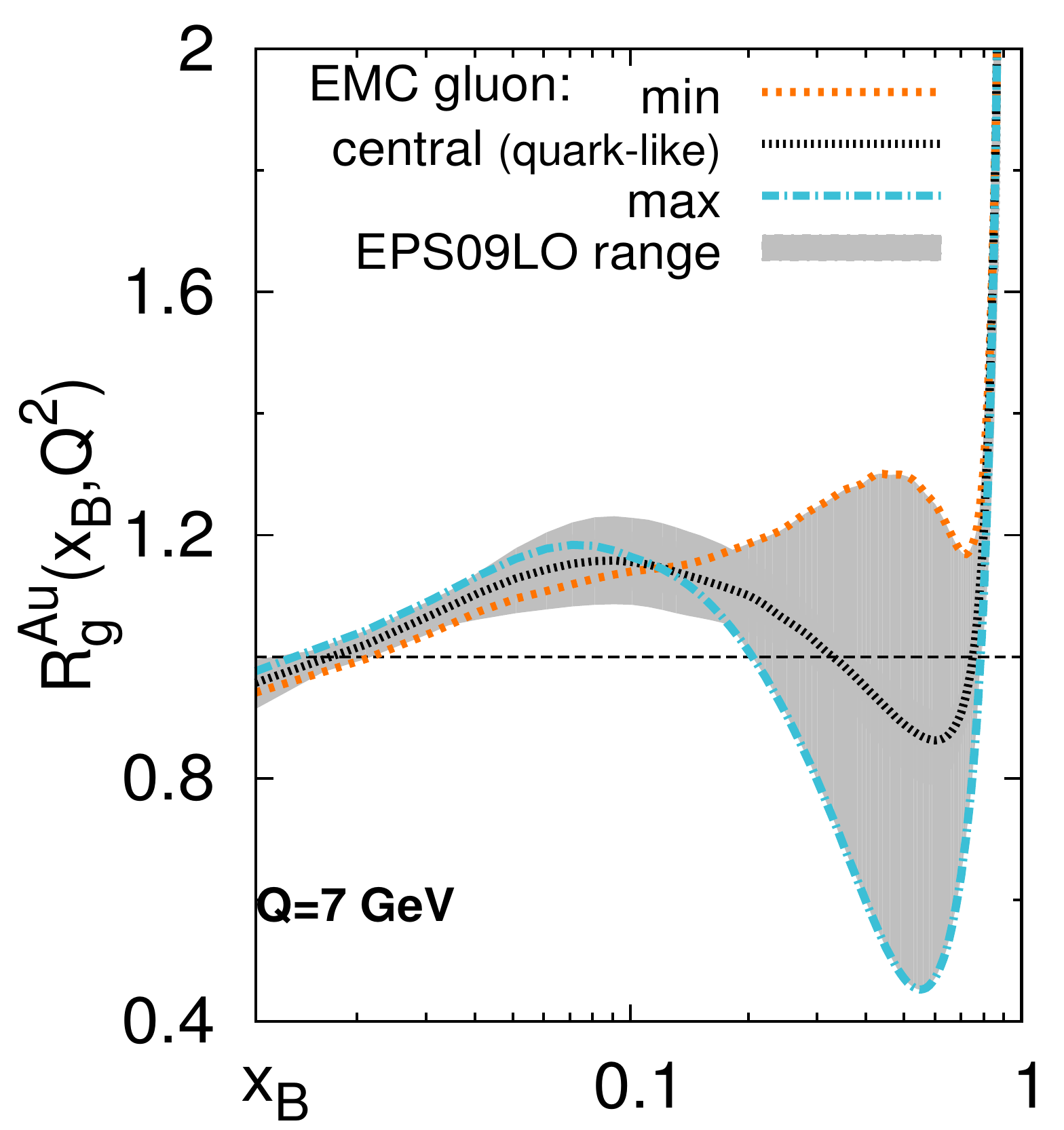}}
%
\subfloat[][Probing the gluon EMC effect with $R^{\Upsilon}_{\dAum}$.]{%
\label{fig:EMC-nPDF-sigma-d}
\includegraphics[width=.49\linewidth]{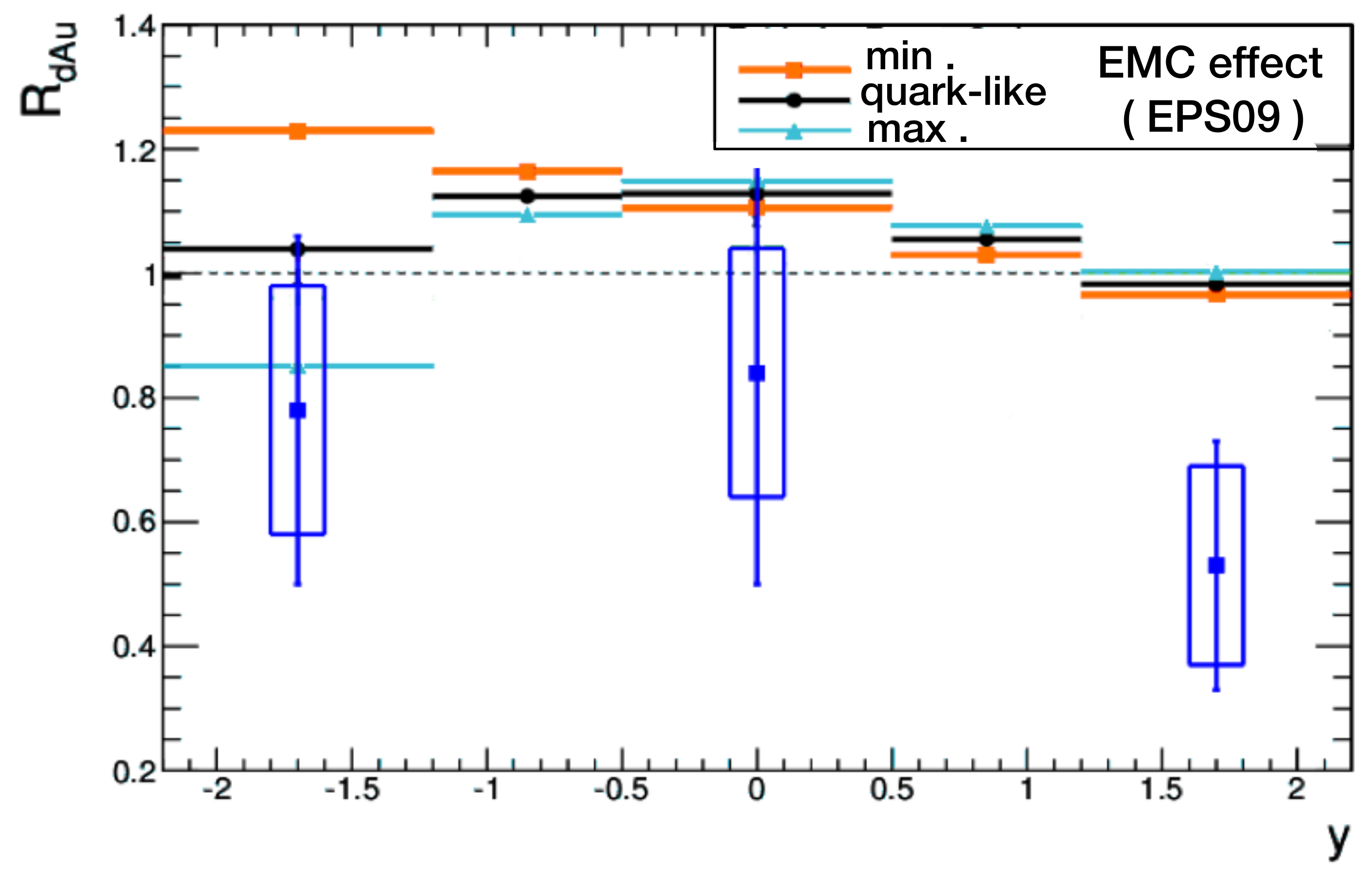}}
\end{center}
\vspace*{-1cm}
\caption{}
\label{fig:EMC-nPDF-sigma}
\end{figure}

To explore 
as widely as possible the impact of any nuclear modification of the gluon PDF, 
we have considered three different parametrisations: EKS98~\cite{Eskola:1998df}, 
EPS08~\cite{Eskola:2008ca} and nDSg~\cite{deFlorian:2003qf} at LO. Yet, they span the current evaluation of the
uncertainty on the gluon nPDF as provided by the newer sets, EPS09~\cite{Eskola:2009uj}, from a small to a very large (anti-)shadowing.
The resulting $R^{\Upsilon}_{\dAum}$ (see the green band in \cf{fig:EMC-nPDF-sigma-b}) has a much flatter rapidity dependence with respect to the data. The forward-$y$ window, with $\left < x_2 \right > \sim 0.008$, corresponds to the shadowing region of the nPDF. However, the expected suppression is not strong enough to match the data. The mid-$y$ window, with $\left < x_2 \right > \sim 0.05$, corresponds to the start of the anti-shadowing region. More precise data is highly desirable to draw any conclsuion on any enhancement of the gluon nPDF in this region, the gluon anti-shadowing existence still being under debate.

To what concerns the gluon EMC effect, we have used three of the EPS09 LO sets: one with a quark-like EMC 
gluon suppression, and the two limiting curves in the region 
$0.35 < x_B < 0.7$ (see \cf{fig:EMC-nPDF-sigma-c}). They translate into the respective $R^{\Upsilon}_{\dAum}$ 
expectations at backward-$y$ shown on \cf{fig:EMC-nPDF-sigma-d}. Despite the large experimental uncertainties, 
the comparison to the data already disfavours 
any nuclear PDF sets that enhance the gluon distribution in this $x_B$ region, while it
favors a sizable gluon EMC suppression, analogous to the quark one or even stronger.

\section{Effective break-up cross-section}
\label{sec:eff-sigma-abs}

The probability for the heavy-quark pair to survive the propagation through the nuclear medium is usually parametrised by an effective 
cross section~$\sigma_{\mathrm{eff}}$. Due to its smaller size, the \bb\ pair should suffer less break-up than the \cc\ pair. Yet, the ratio of their size depends on the evolution stage of the heavy-quark pair. At the production time, this ratio is expected to be $m_b/m_c$. When they are fully formed, it is rather $\frac{\alpha_s(2m_b)}{\alpha_s(2m_c)} \times  \frac{m_b}{m_c}$ as expected from their Bohr radii. 
The relevant timescale for the pair evolution is the formation time. According to the uncertainty principle, it is related to the time needed -- in their rest frame -- to distinguish the energy levels of the $1S$ and $2S$ states 
\cite{ConesadelValle:2011fw}:
$t_f =  \frac{2 M_{b\bar b}}{(M^2_{\mathrm 2S}-M^2_{\mathrm 1S})}  \sim  0.4$~fm for the \upsi.
For our purpose, $t_f$ has to be considered in the rest frame of the target (Au) nucleus. The formation time for different rapidities are given in~\ct{tab:tf}: $t_f$ is significantly larger than the Au radius -- except in the most backward region -- implying that the \bb\ pair is still in a pre-resonant state when traversing the nuclear matter. 

At forward and mid-$y$, this has two implications : 
$\sigma_{\mathrm{eff}}^\Upsilon \sim (\frac{m_c}{m_b})^2 \sigma_{\mathrm{eff}}^{J/\psi } \sim 0.1 \, \sigma_{\mathrm{eff}}^{J/\psi } $, following the 
early-time scaling $m_b/m_c$; and $\sigma_{\mathrm{eff}}$ ought to be the same for 
the \upsi(1S), \upsi(2S) and \upsi(3S) states, since they cannot be distinguished at the time they traverse the nucleus. 
On the contrary, at backward-$y$, we might expect different $\sigma_{\mathrm{eff}}$ for these 3 states for $y<-1$. 
However, the E772 experiment at Fermilab \cite{Alde:1991sw} has measured the 
\upsi(1S) and \upsi(2S+3S) separately at $\sqrt{s}=38.8$ GeV down to negative $x_F$ -- with even smaller formation times -- and 
it observed a similar suppression for the 1S and the $(2{\mathrm S}+3{\mathrm S})$ states. 
Such a result can only be understood if the absorption of the \bb\ resonance is actually very small, preventing us to see any measurable difference 
between the 3 states. In the following, we will consider a range of $\sigma_{\mathrm{eff}}$ from 0 to 1~mb, where 1~mb is a conservative upper bound. In \cf{fig:EMC-nPDF-sigma-b}, this range translates into an uncertainty in $R^{\Upsilon}_{\dAum}$, which is obtained here using EKS98 as the nPDF parametrisation. Even in the strongest absorption case, the resulting $y$-dependence of $R^{\Upsilon}_{\dAum}$ is still not reconciled with the one of the data, which is much steeper at forward-$y$.

\begin{table}[htb!]
\begin{center}\setlength{\arrayrulewidth}{1pt}
\caption{$\Upsilon$ boost and formation time in the gold rest frame as a function of its rapidity
at $\sqrt{s_{NN}}=200$ GeV.}\label{tab:tf}
\begin{tabular}{cccc|cccc}
\hline\hline
 $y$ & $\gamma$ & $t_f$   & \quad& & $y$  & $\gamma$ & $t_f$\\
\hline
-2.0 & 14.4     & 5.8 fm &   &   &  0.0 & 106      & 42 fm\\
-1.5 & 23.7     & 9.5 fm &   &   & +1.5 & 476      & 190 fm\\
-1.0 & 39       & 16 fm &   &   & +2.0 & 786      & 310 fm\\
\hline\hline
\end{tabular}
\end{center}\vspace*{-1cm}
\end{table}

\section{Conclusions and outlook}
\label{sec:conclusions}

We have reported on our recent investigations~\cite{Ferreiro:2011xy} on the \upsi\ production 
in \dAu\ collisions at $\sqrt{s_{NN}}=200$ GeV, for which two RHIC experiments have provided the rapidity dependence. 
Due to the large scale set in by the $\upsi$ mass and the somewhat larger value of $x_2$ compared to the $J/\psi$ case, 
shadowing is found too weak to match the suppression seen at $y>0$. In addition, the data at $y\simeq 0$ does not show 
any excess which would pin down anti-shadowing. Last, but not least, we have argued that the bottomonium survival probability 
to escape the Au nucleus should be quite large, compared to that of the charmonia at the same energy. Even in the strongest 
absorption case, the resulting $y$-dependence of $R^{\Upsilon}_{\dAum}$ cannot be reconciled with that of the data, 
which is more marked at forward-$y$. This discrepancy leaves some room for a fractional parton energy loss for forward 
angle in-medium \upsi\ production, as recently revived in the literature~\cite{Arleo:2010rb} for the \jpsi\ production in \pA\ collisions. 

In the most backward region, the suppression of the $\upsi$ yield may be the first hint of a gluon EMC suppression. 
Moreover, this gluon EMC effect might be stronger than the quark one which still remains to be understood. Better 
precision measurements will be crucial to allow for a quantitative study of the gluon EMC effect, and will offer the 
opportunity to constrain the nuclear gluon distribution at large Bjorken-$x$. Along these lines, it has to be
emphasised that a fixed-target experiment at the LHC, such as the project AFTER, 
would offer $\Upsilon$ yields in proton-nucleus collisions with nuclear targets three orders of magnitude larger than at RHIC
\cite{Brodsky:2012vg,Lansberg:2012kf,proc-QNP}. Such precision studies of $\Upsilon$ production  may in the future provide 
us with fundamental information on the internal dynamics of heavy nuclei such as those studied at RHIC and the LHC.

\vspace*{-0.1cm}




\begin{thebibliography}{99}

\bibitem{Lansberg:2006dh}
  J.~P.~Lansberg, Int.\ J.\ Mod.\ Phys.\ A {\bf 21} (2006) 3857

\bibitem{Brambilla:2010cs}
  N.~Brambilla {\it et al.},
  Eur.\ Phys.\ J.\  C {\bf 71} (2011) 1534.

\bibitem{Rapp:2008tf}
  R.~Rapp, {\it et al.}, 
  Prog.\ Part.\ Nucl.\ Phys.\  {\bf 65} (2010) 209.

\bibitem{Reed:2010zzb}
  R.~Reed {\it et al.}, 
 Nucl.\ Phys.\  A {\bf 855} (2011) 440;
  B.~I.~Abelev {\it et al.},  
  Phys.\ Rev.\  D {\bf 82} (2010) 012004;
%
  K.~B.~Lee, {\it et al.}, 
  PoS {\bf DIS2010} (2010) 077.

\bibitem{Ferreiro:2011xy}
  E.~G.~Ferreiro {\it et al.},
  arXiv:1110.5047 [hep-ph].
  
\bibitem{Glauber:1955qq}
  R.~J.~Glauber,
  Phys.\ Rev.\  {\bf 100 } (1955)  242-248.


\bibitem{Gribov:1968jf}
  V.~N.~Gribov,
  Sov.\ Phys.\ JETP {\bf 29 } (1969)  483.

\bibitem{Gousset:1996xt}
  T.~Gousset, H.~J.~Pirner,
  Phys.\ Lett.\  B {\bf 375} (1996) 349.



\bibitem{Aubert:1983rq}
  J.~J.~Aubert {\it et al.}, 
  Phys.\ Lett.\  B {\bf 123} (1983) 123.
  
\bibitem{Norton:2003cb}
  P.~R.~Norton,
  Rept.\ Prog.\ Phys.\  {\bf 66 } (2003)  1253.
  


\bibitem{Eskola:2009uj}
  K.~J.~Eskola {\it et al.}, 
  JHEP {\bf 0904} (2009) 065.


\bibitem{Arleo:2010rb}
  F.~Arleo, S.~Peigne, T.~Sami, 
  Phys.\ Rev.\  {\bf D83 } (2011)  114036;
%
  F.~Arleo and S.~Peigne,
  arXiv:1204.4609 [hep-ph].
   
\bibitem{Brodsky:1992nq}
  S.~J.~Brodsky and P.~Hoyer,
  Phys.\ Lett.\  B {\bf 298} (1993) 165.

\bibitem{Baier:1996sk}
  R.~Baier {\it et al.}, 
  Nucl.\ Phys.\  B {\bf 484 } (1997)  265.
  
\bibitem{Ferreiro:2008qj}
  E.~G.~Ferreiro {\it et al.}, 
  Eur.\ Phys.\ J.\  C {\bf 61} (2009) 859;
%
  E.~G.~Ferreiro  {\it et al.}, 
  Phys.\ Lett.\  B {\bf 680}, 50 (2009);
%
  {\it Ibid.}, 
  Phys.\ Rev.\  C {\bf 81} (2010) 064911;
  E.~G.~Ferreiro {\it et al.}, 
  Few Body Syst.\  {\bf 53} (2012) 27


  
\bibitem{Vogt:2010aa}
  R.~Vogt,
  Phys.\ Rev.\  C {\bf 81} (2010) 044903.


\bibitem{Brodsky:2009cf}
  S.~J.~Brodsky, J.~P.~Lansberg,
  Phys.\ Rev.\  D {\bf 81} (2010) 051502;
%
  J.~P.~Lansberg,
  Eur.\ Phys.\ J.\ C {\bf 61} (2009) 693;
  {\it Ibid.},
  PoS ICHEP {\bf 2010} (2010) 206.

\bibitem{Klein:2003dj}
  S.~R.~Klein, R.~Vogt,
  Phys.\ Rev.\ Lett.\  {\bf 91} (2003) 142301.


\bibitem{Eskola:1998df}
  K.~J.~Eskola,{\it et al.}, 
  Eur.\ Phys.\ J.\  C {\bf 9} (1999) 61.

\bibitem{Eskola:2008ca}
  K.~J.~Eskola {\it et al.}, 
  JHEP {\bf 0807} (2008) 102.

\bibitem{deFlorian:2003qf}
  D.~de Florian, R.~Sassot,
  Phys.\ Rev.\  D {\bf 69} (2004) 074028.
    
\bibitem{ConesadelValle:2011fw}
  See section 3.1 of Z.~Conesa del Valle, {\it et al.},
  Nucl.\ Phys.\ B\ (PS)  {\bf 214 } (2011)  3.

\bibitem{Alde:1991sw}
  D.~M.~Alde {\it et al.},
  Phys.\ Rev.\ Lett.\  {\bf 66} (1991) 2285.


\bibitem{Brodsky:2012vg}
  S.~J.~Brodsky, F.~Fleuret, C.~Hadjidakis and J.~P.~Lansberg,
  arXiv:1202.6585 [hep-ph].

\bibitem{Lansberg:2012kf}
  J.~P.~Lansberg, S.~J.~Brodsky, F.~Fleuret and C.~Hadjidakis,
  Few Body Syst.\  {\bf 53} (2012) 11

\bibitem{proc-QNP}
  J.~P.~Lansberg, \etal, 
these proceedings.


\end{thebibliography}
\end{document}